# Teacher Module in an Assistance Tool

Adaptating a device to a teaching context and and teacher's preferences

Marilyne Rosselle
MIS Laboratory, UPJV University
MIS-UPJV, 33 rue Saint Leu,
80039 AMIENS CEDEX 1, France
marilyne.rosselle@u-picardie.fr

*Abstract*— **This communication presents the genesis and the implementation of a teacher module, which is included in an Assistance Tool (AT). The teacher module is based on a teacher model for which we did a thorough analysis of the state of the art. The aim of the AT is to help a teacher to design pedagogical devices. Teachers can formulate their needs (assistance in the design) and the AT can relieve them from repetitive tasks related to the deployment of a teaching device (assistance in the deployment).**

## I. INTRODUCTION AND CONTEXT

In TEL (Technology Enhanced Learning), we are used to define student models. But we also need teacher models. Indeed, teachers are asked to make the learner an actor of its learning, and to make learners working with peers (social constructivism). Teachers also have to use Information and Communication Technologies (ICT) for the purpose of learning with their learners. But many teachers feel at ease neither with social constructivism nor with ICT. In particular, the least young ones, who still have numerous years to do and who were not trained to these methods and tools, have difficulty in adapting their practices. Young digital native teachers also meet difficulties. And unfortunately, not all the teachers do have a technical staff to help them.

This research is aimed to help teachers to integrate new methods and new tools in order to adapt their practices and, *in fine*, to help learners. We handle both assistance in designing and assistance in deploying pedagogical devices. For that, D. Leclet and B. Talon have designed an Assistance Tool (AT). M. Hérault and M. Rosselle have designed and implemented the corresponding software. This paper presents the teacher module, a component of the AT. It collects information about teacher to adapt the device proposed to him/her.

This paper is structured in three main sections: we illustrate the context in which the AT can help teachers through an example. Then we present a state of the art concerning teacher models. Afterwards, we detail our module.

## II. EXAMPLE OF THE USE OF THE TEACHER MODULE

To understand how the teacher module can help the teacher, we're describing here, a teaching unit (TU) for which he chooses a pedagogical method and needs to create a device.

**The teaching unit.** Mr Jones is responsible of a "Web programming" TU. The pedagogical staff is composed of 3 teachers. He takes charge of the lectures for about 100 students. For the practical work, the students are divided in 4 groups. All the students received 14 hours of lectures and 26 hours of practical work.

**The pedagogical method.** Mr Jones decided to use the MAETIC pedagogical method. It's a well-defined method ([1]), designed by D. Leclet and B. Talon. MAETIC is based on an active pedagogy that mixes group and project pedagogy in order to allow the learners to learn a domain while realizing a product (*e.g.* a database or flyers). The Domain can be whatever we want. We mainly use it for professional training. We have used it for "Information Systems", "Software Test" or "Communication Supports". MAETIC defines steps that each team of student has to follow. In each step they have to focus on some aspects of the project management and to produce some deliveries related to the defined activities.

**The needed devices.** Students have to create a logbook, in order to (1) inform the teacher of the progress of their work, (2) give the deliveries, and (3) communicate with one another. The logbook can use any technology. This year, we ask learners to use blogs.

Mr Jones has also got a logbook completed by pedagogical resources both for the MAETIC method and the target domain. Moreover he manages a list of the student blogs. He adds a presentation of the pedagogical device and the method he uses. For doing this, he can use any technology. This year, he chose blogs too. The logbook, the pedagogical resources, the links to student blogs, and the presentation constitute his "e-suitcase".

He also uses some specific tools to report absences, evaluate the deliveries, gives feedback or additional information to a specific team or to all the teams. He uses a spreadsheet, a word processor, or some tools that are specific for the target domain. We call all this whole set of tools his "toolbox".

**The added value of the teacher module.** When Mr Jones uses the AT, it collects data, calculates a pedagogical scenario, composes the student teams and generates the student blogs, the teacher's e-suitcase and toolbox (for further information about the AT, see [2] [3]). Within the AT, the teacher module manages data concerning Mr Jones. He describes the TU via some forms (target domain project, client needs for the product, duration of the teaching unit, duration of the sessions for each group of students, pedagogical resources, *etc.*). Moreover he defines his level of knowledge for active pedagogy, group pedagogy, or MAETIC. He chooses his favourite tools for managing his groups. And he chooses the "look & feel" he prefers to assemble his tools. By filling information about him, Mr Jones allows the AT to adapt itself to him in this way:

- Mr Jones has little knowledge about MAETIC. He'd rather process audio information than video information. Thus the AT can propose a spoken presentation of MAETIC.

- Moreover, for practical work, his group is divided in 5 teams the AT generates the 5 students' blogs and it includes the links to these 5 blogs in the e-suitcase.

- And finally, Mr Jones wants to use a spreadsheet, the AT can embed this tool in the toolbox.

### III. STATE OF THE ART

To determine data collected in teacher module, we did a thorough analysis of the state of the art. We were interested in two complementary points of views: education sciences and IT (Information Technology). Thus we structured our presentation by grouping models in education science in Section III.A. Section III.B concerns models in IT. Moreover, as our teacher learns how to introduce new methods or ICT in his teaching strategies, thus we have also studied learner models, even if teacher is not only a learner. That's why the Section III.C is about learner models. Section D collects and synthesises data found in sections A, B and C.

#### A. Teacher Models in Education Sciences

In education sciences, many teacher models exist and are usually richer and more numerous than in IT. They often are under paper format and thus they are not limited by the implementation capacities of computers.

In teacher training, six facets of the teacher's job can describe the teacher model: an instructed master, a technician, an expert craftsman, a reflexive expert, a social actor, and a person. This model is primarily founded on the mapping of three models: four paradigms of Zeichner (1983), four designs of the expertise of Kennedy (1987), and four axes of a formation of Grootaers and Tilman (1991). It leads to the Lang's description of the teacher in six poles (1996): academic, artisanal, applied sciences, professional, person, and social actor.

For [4], the teacher is described in his/her total environment by four groups of variables:

- Predicts (cultural background, formation experience)

- Context (cultural and personal background of the pupils, characteristics of the pupils, dimensions of the class, teaching equipment)

- Process (behaviours of the master, behaviours of the pupils, behaviour changes observed on the pupils)

- Product (the development of the pupils)

Another often quoted model is the Paquay's one (1994). It identifies four models of the teaching profession: magister teacher, technician teacher, engineer teacher, and professional teacher. Brousseau (1995) proposes to reconsider the modelling of the teacher in the didactic theory: "there isn't strictly speaking a teacher model in the theory of the situations. We must thus seek to identify the `realities' which relate to him/her with the disturbances and the regulations that he/she produces and ensures in the didactic system".

#### B. Teacher Models in Information Technology

All work in IT related to a teacher model only took place in education specific domains where IT acts as a supporting technology. In TEL the teacher model is often limited to a model of the knowledge of the teacher on the domain taught: the domain model or the model of the expert. Instead, we are interested in the teaching knowledge. But little work exists on the teaching knowledge.

We found elements on the tutoring role of the teacher (especially in Intelligent Tutoring Systems [5][6]), but not on his/her strategies, teaching choices and technological preferences. However, there is some teaching knowledge in the tutoring module of Wheeldon [7]. It is used to determine the comprehensive strategies (*e.g.* order of presentation of materials) and the local tactics (*e.g.* to guide the user to carry out an action). However, if Wheeldon explains well what the tutoring module does, he doesn't present a model of this module. In [8], the teacher model contains the teacher's needs, the functionalities of tools he/she prefers and data on his/her physical environment. [9] uses a "pedagogue profile" in his educational studio. For each module of this studio, there is a model of interface corresponding to the pedagogue profile: simple user, advanced user, or super user. These profiles correspond to rights in the database of the system but are not described more finely. In [10], the teacher is described according to three components: the domain, the teaching style and the thinking style. Another model proposes modelling teacher role: author, designer, resource organizer, scenario writer [11].

#### C. Learner Models in TEL

A TEL can be based on a learner model. Many learner models in the TEL community exist. [12] defines a learner profile as a set of data characterizing his/her knowledge, competences, conceptions, or his behaviour. For [13] a learner profile informs about his/her knowledge, control level and possible relations. Learner models in IT have been standardised, especially in the context of use of a platform (PAPI [14], IMSRDCEO [15], and IMSLIP [16]). The aim was to facilitate the storage and exchange of data to provide a help to the management of the education institutions. However information is not described sufficiently precisely in them. In [17], authors propose an object model where pieces of information on the learner are grouped in classes. In [18], authors present two overlay learner models, *i.e.* one regards the acquired knowledge as a percentage from the target knowledge. In [19], authors use a learner model to store objectives, preferences and knowledge. The Felder-Silverman model [20] classifies learners according to the relevant ways that they use to perceive information and to process data. Thus, learners are identified by their type: sensory or intuitive, visual or verbal, inductive or deductive, active or reflexive, sequential or global.

#### D. Toward a synthetical view...

We found very interesting elements for the teacher model of the AT from this literature. Unfortunately, there were no room enough to explain each item we found in Section A, B, and C. Thus we are now trying to organise them synthetically.

We can gather items found in previous sections in dimensions as [21] began it (the dimensions that we have added are in italic). The dimensions concern:

- Personal data (*e.g.* name, surname, email, password),
- Skills (*e.g.* organize, negotiate, coach, motivate)
- Knowledge (*e.g.* domain knowledge, pedagogical knowledge, technical knowledge),
- Behaviour (*e.g.* behaviour of the teacher when managing groups),
- Personality traits (psychological data)
- Experience (*e.g.* diploma, formation, specialisation,)
- Working context (*e.g.* cultural background, type of school, dimension of the class, teaching equipment),
- Centre of interest (*e.g.* sciences),
- Preferences (*e.g.* tools that he is used to),
- *Personality types* (*e.g.* prefer to process visual information),
- *Social relationship* (*e.g.* colleagues with whom he collaborates).
- *Teaching style* (e.g. guide on the side or sage on the stage, slides or chalk and talk), and
- *Claimed teaching theory* (e.g. constructivism)

All these dimensions provide a generic multi-dimensional model that can be used in various contexts. Indeed, one can choose the dimensions that are relevant for its context or car implement all these dimensions. A full use of all these dimensions in our AT would have been interesting. However, in a pragmatic approach, we have chosen to only retain, in a first version, elements that facilitate the adaptation of the AT or elements that favour the production of a suitable device.

In this state of the art, teacher models are most often formalized on paper (*e.g.* in a form). When a model for computers exists, teacher models are mostly presented either by a UML (Unified Modelling Language) class diagram or by an ER (Entity-Relationship) model, or expressed in MOT (Modelling by Typified Objects) formalism. Data are grouped semantically into entities or classes (*e.g.* there is an entity or a class for the teacher's knowledge).

The implementations are either in object-oriented format (case of a UML models), or directed towards the production of a database (case of the entity-relationship models), or take the form of ontologies (case of MOT models).

IV. THE TEACHER MODULE

The AT aims to adapt its behaviour and produce a device customised for each teacher using the teacher module data. The following section describes the modelling involved in the module. Then we detail the implementation. Finally, we show how the teacher module is used in the context of our initial example.

*A. Modelling*

Even if it seems natural to separate data on teacher and teaching, we have seen in Section III that, in numerous models (*e.g.* [8][18]), data on the teacher (preferences, psychological profile, *etc.*) and those on the teaching unit (duration of the unit, number of learners, *etc.*) are put in the same space. We separate data on the teacher and those on the teaching unit because, in the context of the AT use, a teacher can handle several teaching units. By separating data, we avoid redundancies. Thus, the teacher module leans on a teacher model and a teaching unit model (what is to be taught, in which context, etc.). The teaching unit model will be presented in another publication.

Fig. 1 presents the dimensions that we retain. To represent them we chose a UML class diagram partly because the AT is implemented in Java. We have also chosen it because UML seems to be better known that the other formalism.

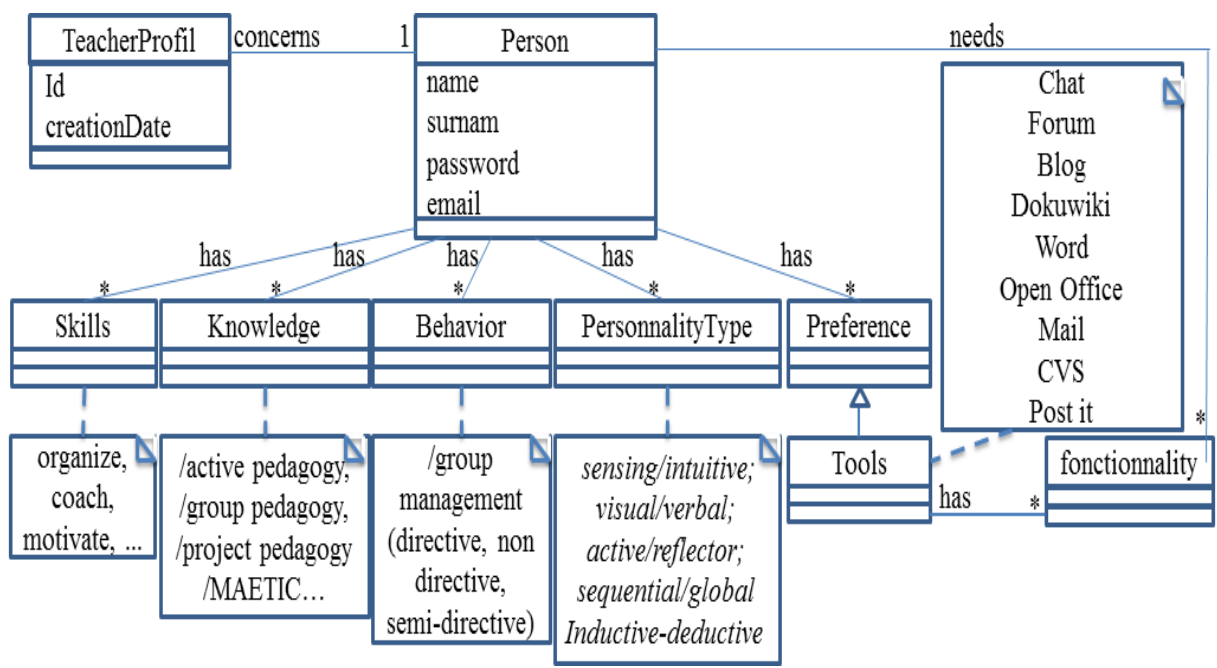

Figure 1. Our teacher model - UML class diagram

In Fig. 1, we can see that we retain data on the teachers' skills, knowledge, and behaviour, functionalities of the tools he/she's waiting for, and his/her personality type (which include the selected psychological type). In this diagram, classes are types of data gathered while notes give examples of stored data. This model has already been published in French in [22].

The model reifies in a teacher profile (the model is generic while a profile is specific to a teacher). For example, Mr Jones is a teacher who has competences for coaching and motivating. He knows active pedagogy. He manages groups in a directive way. He is an intuitive, verbal, deductive, active, and sequential teacher. He's used to the chat, Word, Excel, and wishes to use a functionality of data storage.

*B. Implementation*

The implementation language of the AT is Java (version 1.6). The architecture choices for the teacher module depend partly on the choices made for the AT.

*1) The general architecture*

Because our end users can have difficulties with ICT and IT, we wanted to have them installed to the fewer things as possible. Therefore, we chose to have them used the AT via their browser. We opted for a client-server architecture. The logic of the AT is on the server side. Teacher accesses it via their browser on the client side. The link between both client and server sides is made through an RPC (Remote Procedure Call) communication provided via the Internet. Two communication flows are necessary. The first one is the client flow to the server. It is used when some actions on the interface are producing calls to services. Services are started on the server side. The second asynchronous flow receives the server answers. It is compatible with Tomcat (or any application server that implements communicating through JSONP or GWT RPC).

*2) Added choice for the graphical interface*

For the realisation of the graphical interface, we opted for the GWT technology (Google Web Toolkit, version 2.4). GWT

*3) Added choice for the management of the teacher model*

We have to save teacher data somewhere. We could have used a file, or a database. But we had already known that we would have needed to make some inferences from some data to generate other data later. Ontology is exactly a way especially designed to reach this purpose. This choice also allows us to implement the expert knowledge in rules that are outside the business logic of the AT, and hence are more easily maintainable (*e.g.* these rules prescribe interesting activities or propose adapted tools). Thus we have chosen to use an ontology and to populate it, with the data, which are collected by the teacher module. To represent our ontology, we chose OWL-DL, which is a language that is based on the formalism of the logic of description. To do that, we used the "Protégé" OWL editor. And to be able to manage our ontology with Java, we chose to use the JENA API.

Nevertheless, we use a database, for the login phase. It only stores the few data about the "person", that are used when the teacher registers himself/herself or when he/she logs in. *Thus we added a database management system (DBMS) with the application server*. We have chosen MySQL (implemented in easyphp or mamp, whether we tested our application on a PC computer or on a Mac computer). Moreover, this data separation allows us to assure anonymity in the ontology, which only stores the database identifier for the teacher.

*4) Architecture summary*

In summary, for the architecture, we have on the server side: an application server, a DBMS, an ontology, the Jena API, the business logic and the GWT graphical interface. On the client side, we have the graphical interface in JavaScript that is tailored to the browser. The teacher interacts with the browser. All these elements are shown in Fig. 3. On Fig. 2, we see the server side on the left; the client side is on the right. They communicate through the Internet by using RPC.

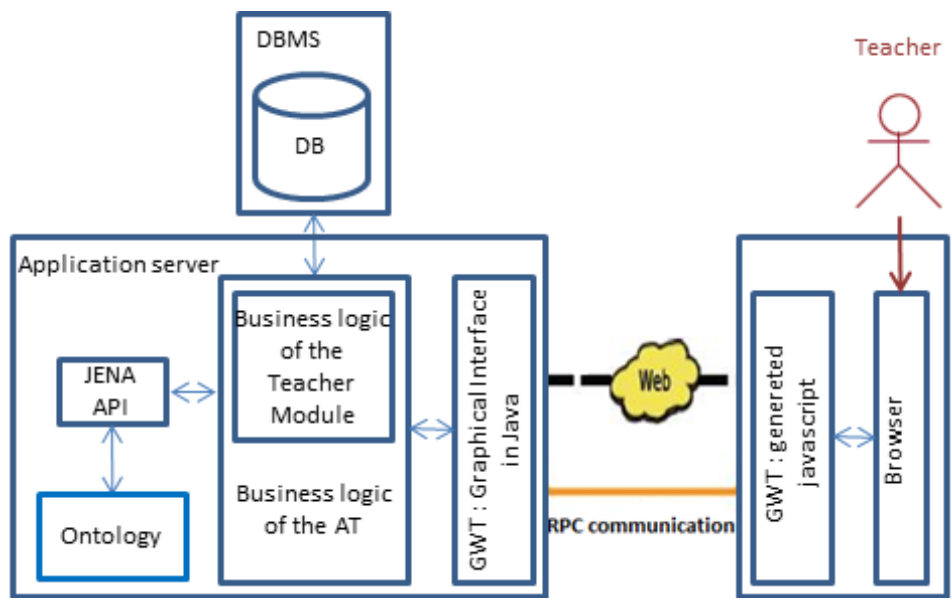

Figure 2. Architecture of the teacher module

allows defining the interactions between the AT (the business logic) and the client interface in Java. This code is compiled in Javascript and executed on the client side in a browser. GWT insures portability with 6 popular web browsers by generating the necessary JavaScript/HTML/DOM codes for each one. Moreover, the GWT technology is compatible with several application servers.

*5) Structure of the teacher module*

Inside the teacher module, there are the client package and the server package. In that client package, we have the TeacherModule class. It uses classes assigned to:

- Either in the GraphicalElem package (*e.g.* the TeacherForm class, which defines the form to collect the teacher profile),

- Or in the BusinessElem package (*e.g.* the User class which is used to identified the user in the application, and the TeacherProfil, which memorised the data collected about the teacher),
- Or in the Services package (the classes which are linked with the Server Package).

In our server package, we have classes that implement the services. Here they store data from the teacher profile into ontology file. In practice, we have a folder (named users) for storing files about our users. In this folder we have a folder for each user (*e.g.* User1, User2). In each folder, we have a "profil.owl" file when the user has already defined his/her teacher profile. This folder is also used to store others files about this user (*e.g.* the teaching units he/she has entered).

### C. Using the teacher module in the assistance tool within the initial example

Here is an illustration of the role of the teacher module. First Mr Jones logs in the AT. Then he can fill his profile (that is handled by the teacher module). If he doesn't want to fill the form, the AT has a standard behaviour and the produced device is a standard one. Mr Jones fill is profile as shown in Fig. 3.

according to information collected in a state of the art done in both education sciences and IT. We propose a teacher module, which selects collected data. We also designed a learning unit model. The proposed models support the production of a device adapted to the teacher. This production is based on inferences rules collected from experts. The rules are applied to data collected by the teacher module and stored in an ontology.

In an early stage, the teacher and the teaching unit models led to the setting of two paper forms presented to teachers for collecting data. The first evaluation of the forms was made in 2009 in order to know whether the teacher can fill it well or not. Then, a first implementation of these forms was made in 2010. The forms were developed in Java & GWT and data were stored in an ontology. This implementation allows us to validate the model in a second way, *i.e.* it is sufficient to be used by inference motor in order to generate a pedagogical scenario.

A first version of The AT, which includes a complete Teacher Module, has been stabilised and used in October 2011. It has video presentations for active pedagogy, group pedagogy and MAETIC. But these video should be improved. Now, the AT is still under development. The part which is under development concerns the generation of the blogs needed by

Figure 3. The first part of the teacher profile form.

When Mr Jones fills the profile, data is stored in a "profil.owl" file and the behaviour of the AT is customized. Second the AT consults the teacher module to select the way it will present the needed information: it will not present active, group and project pedagogies because Mr Jones indicates that he knows them. But it will present the MAETIC method by using audio information and ordering it in order to favour deduction.

### V. CONCLUSION

In this article, we explained how we help a teacher to deploy a new pedagogical method and device that is adapted to this method and to him/her. To that end, we are developing an Assistance Tool (AT). The AT contains a teacher module. This module is based on a teacher model. The model is established teacher and learners. However the teacher module is completed. The first version of the generated e-suitcase and toolbox for the teacher is almost finished. Indeed we work incrementally. Therefore, it doesn't allow the teacher to choose between several look & feel and to choose additional tools. A presentation (and a validation if possible) of the models and of the AT, in an ecological environment, is planned at the University Fez in Morocco in spring 2012.

Another trend concerns the psychological part of the teacher model. In the current system, the teacher fills in data about himself/herself (*e.g.* he/she says whether he/she is verbal or visual). In the future, we propose to refer to the 44 questions of the Felder-Silverman [20] quiz in order to automatically deduce the psychological type of the teacher.

Finally, we would evaluate the generic feature of the multi-dimensional model by applying it in other contexts (colleagues work on a platform that aims to help new teachers, other ones work on a social practice community for teachers).

ACKNOWLEDGMENTS

We want to thank Dominique Leclet and Bénédicte Talon, who have designed the AT, and Mathieu Hérault, who has developed the main part of the first version of the teacher module and of the assistance tool with my modest contribution.

REFERENCES

[1] D. Leclet and B. Talon, "Binding the gap between professional context and university: e-suitcase MAETIC for a real world experience", Interactive Computer aided Learning, 2008.

[2] D. Leclet and B. Talon, "Assessment of a method for designing e-learning devices", World Conference on Educational Multimedia, Hypermedia and Telecommunications, ED-MEDIA, AACE/Springer-Verlag, pp. 1-8, 2008.

[3] D. Leclet and M. Rosselle, "Designing pedagogical devices - Assistance tool for teachers", EDUCON, in press.

[4] B.D. Ndiaye, "Étude des conceptions des enseignants du Sénégal sur le métier, en référence au modèle de l'enseignant-professionnel", Phd thesis of UCL (Université catholique de Louvain), Faculté de Psychologie et des Sciences de l'Education, Belgique, 2003.

[5] J. Bourdeau and M. Grandbastien, "Modeling tutoring knowledge" in *Advances in Intelligent Tutoring Systems*, R. Nkambou, R. Mizoguchi, and J. Bourdeau, Eds. , 2010, 123-143.

[6] IJAIED special issues on Authoring ITS, vol 19, 2009.

[7] A. Wheeldon, "Improving human computer interaction in intelligent tutoring system" http://eprints.qut.edu.au/16587/

[8] M.-N. Bessagnet and D. Hérin, "Une approche pour l'aide à l'usage des technologies de l'information et de la communication pour les enseignants du supérieur" Environnements Informatiques pour l'Apprentissage Humain, ATIEF & INRP, pp. 523–526, 2003.

[9] C. Marquesuzaà, "OMAGE : outils et méthode pour la spécification des connaissances au sein d'un atelier de génie éducatif". Ph'd thesis of the University of Pau (France), 1998.

[10] K. Wang and P. Trigano, "eCH, a course help tool for teacher", 9th international conference on web based learning, ICWL 2010, Springer LNCS, 2010.

[11] E. Villiot-Leclercq, "conception de scénarios pédagogiques : un dispositif d'assistance pour soutenir l'interaction entre l'enseignant et l'environnement ExploraGraph", Scénarios, J.-P. Pernin and H. Godinet, Eds., pp. 83-88, 2006.

[12] C. Eyssautier-Bavay, S. Jean-Daubias, and J.-P. Pernin, "A model of learners profiles management process", Artificial Intelligence in Education, V. Dimitrova, R. Mizoguchi, B. du Boulay, and A. Graesser, Eds., Brighton (GB). Frontiers in Artificial Intelligence and Applications 200, IOS Press, ISBN 978-1-60750-028-5, pp. 265-272, 2009.

[13] C. Eyssautier-Bavay, S. Jean-Daubias, and J.-P. Pernin, "A model of learners profiles management process". V. Dimitrova, R. Mizoguchi, B. du Boulay, and A Graesser, Eds., Artificial Intelligence in Education, AIED, IOS Press, pp. 265–272, 2009.

[14] Public and Private Information for Learners (PAPI Learner) http://metadata-standards.org/Document-library/Meeting-reports/SC32WG2/2002-05-Seoul/

[15] Reusable Definition of Competency or Educational objective (IMS RDCEO), http://www.imsglobal.org/competencies/

[16] Learner Information Package (IMS-LIP) http://www.imsglobal.org/profiles/lipbest01.html

[17] M. T. Ramandalahy, P. Vidal, and J. Broisin, "opening learner profiles across heterogeneous applications". ICALT, IEEE Computer Society, pp. 504–508, 2009.

[18] S. Bull, P. Gardner, N. Ahmad, J. Ting, and B. Clarke, "Use and trust of simple independent open learner models to support learning within and across courses". UMAP, G.-J. Houben et al., Eds. 2009. LNCS, vol. 5535, Springer-Verlag, Heidelberg, pp. 42–53, 2009.

[19] E. García, C. Romero, S. Ventura, and C. de Castro, "Evaluating web based instructional models using association rule mining", UMAP, G.-J. Houben G.-J. et al., Eds. LNCS, vol. 5535, Springer-Verlag, Heidelberg, pp. 16–29, 2009.

[20] R.M. Felder and L.K. Silverman, "Learning and Teaching Styles in Engineering Education." Engr. Education, 78(7), 674-681, 1988.

[21] T. Condamines, "A teacher is also a learner: toward a teacher multidimensional model", Conference on Educational Multimedia, Hypermedia & Telecommunications, EdMedia, Lisbonne, Portugal pp. 1727-1732, 2011.

[22] M. Rosselle, D. Leclet, and B. Talon, "Le Modèle de l'enseignant d'un atelier de génie pédagogique pour la conception de dispositif pédagogique", Technologies de l'Information et de la Communication pour l'Enseignement, TICE'2010, Nancy-Metz, France, pp1-8 (electronic publication), december 2010.